\documentclass[aps,prd,preprintnumbers,amsmath,amssymb,nofootinbib,11pt]{revtex4}
\usepackage{eepic}
\usepackage{indentfirst}
\usepackage{mathrsfs}
\usepackage{fancyhdr}
\usepackage{ulem}
\usepackage{float}
\usepackage{graphicx}
\usepackage[
colorlinks=true,
linkcolor=black,
breaklinks=true,
urlcolor=blue,
citecolor=green]{hyperref}
\usepackage{epstopdf}

\newcommand{\be}{\begin{equation}}
\newcommand{\ee}{\end{equation}}
\newcommand{\ba}{\begin{array}{c}}
\newcommand{\ea}{\end{array}}
\renewcommand{\L}{\mathscr{L}}

\newcommand{\bra}{\langle}
\newcommand{\ket}{\rangle}
\newcommand{\nn}{\nonumber}
\newcommand{\MeV}{\,\text{MeV}}

\renewcommand{\vec}[1]{\mathbf{#1}}
\newcommand{\diff }{{\text{d}}}

\begin{document}

\title{\boldmath Chromopolarizability of charmonium and $\pi\pi$ final state interaction revisited   }

\author{ Yun-Hua~Chen}\email{yhchen@ustb.edu.cn}
\affiliation{School of Mathematics and Physics, University
of Science and Technology Beijing, Beijing 100083, China}

\begin{abstract}

The chromopolarizability of a quarknonium describes the quarknonium's interaction with soft gluonic fields
and can be measured in the heavy quarkonium decays. Within the framework of dispersion theory which consider the $\pi\pi$ final state interaction (FSI) model-independently, we analyze the transition
$\psi^\prime\to J/\psi\pi^+\pi^-$ and obtain the chromopolarizability $\alpha_{\psi^\prime \psi}$ and the parameter $\kappa$.
It is found that the $\pi\pi$ FSI plays an important role in extracting the chromopolarizability from the experimental data.
The obtained chromopolarizability with the FSI is reduced to about 1/2 of that without the FSI. With the FSI, we determine the chromopolarizability $|\alpha_{\psi^\prime\psi}|=(1.44\pm 0.02)$ GeV$^{-3}$ and the parameter $\kappa=0.139\pm 0.005.$
Our results could be useful in studying the interactions of charmonium with light hadrons.

\end{abstract}


\maketitle

\newpage

\section{Introduction}

The chromopolarizability $\alpha$ of a quarkonium parametrizes the quarkonium's effective interaction with soft gluons, and it is
an important quantity in the heavy quark effective theory.
Within the multipole expansion in QCD in terms of the chromopolarizability, many processes can be described, including
the hadronic transtions between quarkonium resonances~\cite{Voloshin:1980zf,Novikov:1980fa} and the interaction of slow quarkonium
with a nuclear medium~\cite{Sibirtsev:2005ex}. A recent interest
of the chromopolarizabilities of $J/\psi$ and $\psi^\prime$ comes from the hadrocharmonium~\cite{Voloshin:2007dx,Dubynskiy:2008mq,Sibirtsev:2005ex,Eides:2015dtr,Tsushima:2011kh} interpretation of the
$P_c^+(4380)$ and $P_c^+(4450)$ observed by the LHCb Collaboration, and it is found that
the $P_c^+(4450)$ can be interpreted as a $\psi^\prime$-nucleon bound state if $\alpha_{\psi^\prime}/\alpha_{J/\psi}\simeq 15$~\cite{Polyakov:2018aey}.

There are a few studies of the chromopolarizabilities of $J/\psi$ and $\psi^\prime$, some of which are not in line with each others.
Calculated in the large-$N_c$ limit in the heavy quark approximation, the values of the chromopolarizabilities of the $J/\psi$
and $\psi^\prime$ are obtained: $\alpha_{J/\psi}\approx 0.2$ GeV$^{-3}$ and $\alpha_{\psi^\prime}\approx 12$ GeV$^{-3}$~\cite{Peskin:1979va,Eides:2015dtr}.
Within a quarkonium-nucleon effective field theory, the chromopolarizability of the $J/\psi$ is determined
through fitting the lattice QCD data~\cite{Kawanai:2010ev} of the $J/\psi$-nucleon potential, and the result is $\alpha_{J/\psi}=0.24$ GeV$^{-3}$ ~\cite{Brambilla:2015rqa,TarrusCastella:2018php}. While based on an effective potential formalism given in Ref.~\cite{Voloshin:1979uv} and a recent
lattice QCD calculation~\cite{Sugiura:2017vks}, the chromopolarizabilities of $J/\psi$ is extracted to be  $\alpha_{J/\psi}=(1.6\pm0.8)$ GeV$^{-3}$~\cite{Polyakov:2018aey}.  On the other hand, the determination of the transitional chromopolarizability $\alpha_{\psi^\prime\psi}\equiv\alpha_{\psi^\prime\to J/\psi}$ is of importance since it
acts a reference benchmark for either of the diagonal terms due to the Schwartz inequality: $\alpha_{J/\psi}\alpha_{\psi^\prime}\geq \alpha_{\psi^\prime\psi}^2$~\cite{Voloshin:2007dx}. The perturbative prediction in the large $N_c$ limit is $\alpha_{\psi^\prime\psi} \approx -0.6$ GeV$^{-3}$~\cite{Peskin:1979va,Eides:2015dtr}. While extracted from the process of $\psi^\prime \to J/\psi \pi\pi$, the result is
$|\alpha_{\psi^\prime\psi}| \approx 2$ GeV$^{-3}$~\cite{Voloshin:2004un,Voloshin:2007dx}. Taking account of the $\pi\pi$ FSI in a chiral unitary approach, it is found that the value of $|\alpha_{\psi^\prime\psi}|$ may be reduced to about $1/3$ of that without the $\pi\pi$ FSI~\cite{Guo:2006ya}.

Since the FSI plays an important role in the heavy quarkonium transitions and modifies the value of $\alpha_{\psi^\prime\psi}$ significantly, it is thus necessary to account for the FSI properly. In this work we will use the dispersion theory to take into account of the $\pi\pi$ FSI and extract the value of $\alpha_{\psi^\prime\psi}$. Instead of the chiral unitary approach~\cite{Oller:1997ti,Guo:2006ya}, in which the scalar mesons ($\sigma$, $f_0(980)$, and $a_0(980)$) are dynamically generated, in the dispersion theory the $\pi\pi$ FSI is treated in a model-independent way consistent with $\pi\pi$ scattering data.
Another update of our calculation is that we consider the FSIs of separate partial waves, namely the $S$- and $D$-waves, instead of only accounting
for the $S$-wave as in the parametrization in Refs~\cite{Voloshin:2004un,Guo:2006ya}.

The theoretical framework is described in
detail in Sec.~\ref{theor}. In Sec.~\ref{pheno}, we fit the decay
amplitudes to the data for the $\psi^\prime\to J/\psi\pi^+\pi^-$ transition, and determine the
the chromopolarizability $\alpha_{\psi^\prime \psi}$ and the parameter $\kappa$. A brief summary will be
presented in Sec.~\ref{conclu}.

\section{Theoretical framework}
\label{theor}

First we define the the Mandelstam variables of the decay process $\psi^\prime(p_a) \to J/\psi(p_b) \pi(p_c)\pi(p_d)$
\begin{align}
s &= (p_c+p_d)^2 , \qquad
t=(p_a-p_c)^2\,, \qquad u=(p_a-p_d)^2\,.
\end{align}

The amplitude for the $\pi^+\pi^-$ transition between $S$-wave states $A$ and $B$ of heavy quarkonium
can be written as~\cite{Voloshin:2007dx,Brambilla:2015rqa}
\begin{equation}\label{eq.MultipoleAmplitude}
M_{AB}=2\sqrt{m_Am_B}\alpha_{AB}\langle
\pi^+(p_c)\pi^-(p_d)|\frac{1}{2}\vec{E}^a\cdot \vec{E}^a|0\rangle=\frac{8\pi^2}{b}\sqrt{m_A m_B}\alpha_{AB}(\kappa_1 p^0_c p^0_d-\kappa_2 p^i_c p^i_d),
\end{equation}
where the factor $2\sqrt{m_Am_B}$ appears due to the relativistic
normalization of the decay amplitude, $\alpha_{AB}$ is the chromopolarizability, and $\vec{E}^a$ denotes
the chromoelectric field. $b$ is the first coefficient of the QCD beta function,
$ b=\frac{11}{3}N_c-\frac{2}{3}N_f,$ where $N_c=3$ and $N_f=3$ are the number of colors and of light flavors, respectively. $\kappa_1=2-9\kappa/2$, and $\kappa_2=2+3\kappa/2$, where $\kappa$ is a parameter that can be determined from the data.

The above result of the QCD multipole expansion together with the
soft-pion theorem can be reproduced by constructing a chiral
effective Lagrangian for the $\psi^\prime \to J/\psi \pi\pi$
transition. Since the spin-dependent interactions are suppressed by the charm mass, the heavy quarkonia
can be expressed in term of spin multiplets, and one has $J \equiv \vec{\psi}
\cdot \boldsymbol{\sigma}+\eta_c$, where $\boldsymbol{\sigma}$
contains the Pauli matrices and $\vec\psi$ and $\eta_c$
annihilate the $\psi$ and $\eta_c$ states, respectively~\cite{Guo2011}. The effective
Lagrangian, at the leading order in the chiral as well as the heavy-quark
nonrelativistic expansion, reads~\cite{Mannel,Chen2016,Chen:2016mjn}
\begin{equation}\label{LagrangianPsiPrimePsipipi}
\L_{\psi\psi^{\prime}\pi\pi}
= \frac{c_1}{2}\bra J^\dagger J^\prime \ket \bra u_\mu u^\mu\ket
+\frac{c_2}{2}\bra J^\dagger J^\prime \ket \bra u_\mu u_\nu\ket v^\mu v^\nu
+\mathrm{h.c.} \,,
\end{equation}
where $v^\mu=(1,\vec{0})$ is the velocity of the heavy quark.
The Goldstone bosons of the spontaneous breaking of
chiral symmetry can be parametrized according to
\be
u_\mu = i \left( u^\dagger \partial_\mu u\, -\, u \partial_\mu u^\dagger\right) \,, \qquad
u^2 = e^{i {\Phi}/{ F_\pi}}\,, \qquad
\Phi =
\begin{pmatrix}
   \pi ^0  & \sqrt2{\pi^+ }  \\
   \sqrt2{\pi^- } & -\pi ^0  \\
\end{pmatrix} ,
\ee
where $F_\pi=92.2\MeV$ denotes the pion decay constant.

The amplitude obtained by using the effective Lagrangians
in Eq.~\eqref{LagrangianPsiPrimePsipipi} is
\begin{equation}
\label{eq.ChiralAmplitude}
M(s,t,u)
= -\frac{4}{F_\pi^2}( c_1 p_c\cdot p_d +c_2 p_c^0 p_d^0)\,.
\end{equation}
Matching the amplitude in Eq.~\eqref{eq.MultipoleAmplitude} to that in Eq.~\eqref{eq.ChiralAmplitude}, we
can express the low-energy couplings in the chiral
effective Lagrangian in terms of the chromopolarizability $\alpha_{AB}$ and the parameter $\kappa$
\begin{align}
c_1&= -\frac{\pi^2\sqrt{m_{\psi^\prime}m_\psi}F_\pi^2}{b}\alpha_{\psi^\prime\psi}(4+3\kappa) , \nn\\
c_2&= \frac{12\pi^2\sqrt{m_{\psi^\prime}m_\psi}F_\pi^2}{b}\alpha_{\psi^\prime\psi}\kappa\,.\label{eq.Matching}
\end{align}

The partial-wave decomposition of $M(s,t,u)$ can be easily performed by using
the relation
\begin{equation}
p_c^0 p_d^0 =
\frac{1}{4} \left(s+\vec{q}^2\right)-
\frac{1}{4}\vec{q}^2 \sigma_\pi^2 \cos^2\theta \,,
\end{equation}
where $\vec{q}$ is the 3-momentum of the final vector meson in
the rest frame of the initial state with $
|\vec{q}|=
\big\{[(m_{\psi^\prime}+m_\psi)^2-s]/[(m_{\psi^\prime}-m_\psi)^2-s]\big\}^{\frac{1}{2}}/(2m_{\psi^\prime}) $,
$\sigma_\pi \equiv \sqrt{1-4m_\pi^2/s}$, and $\theta$
is the
angle between the 3-momentum of the $\pi^+$ in the rest frame of the $\pi\pi$
system and that of the $\pi\pi$ system in the rest frame of the initial
$\psi^\prime$.

Parity and $C$-parity conservations require the pion pair to have
even relative angular momentum $l$. We only consider the $S$- and $D$-wave
components in this study, neglecting the effects of higher
partial waves. Explicitly, the $S$- and $D$-wave components of
the amplitude read
\begin{align}
M_0^{\chi}(s)&= -\frac{2}{F_\pi^2}
\bigg\{c_1 \left(s-2m_\pi^2 \right)
+\frac{c_2}{2} \bigg[s+\vec{q}^2\Big(1  -\frac{\sigma_\pi^2}{3} \Big)\bigg]\bigg\} ,\nn \\
M_2^{\chi}(s)&=\frac{2}{3 F_\pi^2}c_2 \vec{q}^2 \sigma_\pi^2\,.\label{eq.AmplitudeChiral}
\end{align}

There are strong FSI in the $\pi\pi$ system especially in the
isospin-$0$ $S$-wave, which can be taken into account
model-independently using dispersion theory~\cite{Kang,Isken:2017dkw,KubisPlenter,Ropertz:2018stk,Chen2016,Chen:2016mjn,DLY-MRP14}. We will use the Omn\`es solution to obtain the amplitude
including FSI. In the region of elastic
$\pi\pi$ rescattering, the partial-wave unitarity conditions
read
\begin{equation}\label{eq.unitarity1channel}
\textrm{Im}\, M_l(s)= M_l(s)
\sin\delta_l^0(s) e^{-i\delta_l^0(s)}\,.
\end{equation}
Below the inelastic threshold, the phases $\delta_l^I$ of the partial-wave amplitudes
of isospin $I$ and angular momentum $l$ coincide with the
$\pi\pi$ elastic phase shifts modulo $n\pi$, as required by Watson's
theorem~\cite{Watson1,Watson2}.
It is known that the standard Omn\`es solution of Eq.~\eqref{eq.unitarity1channel},
\be\label{OmnesSolution}
M_l(s)=P_l^{n}(s)\Omega_l^0(s) \,, \ee where
the $P_l^{n}(s)$ is a polynomial, and the Omn\`es function is defined
as~\cite{Omnes}
\begin{equation}\label{Omnesrepresentation}
\Omega_l^I(s)=\exp
\bigg\{\frac{s}{\pi}\int^\infty_{4m_\pi^2}\frac{\diff x}{x}
\frac{\delta_l^I(x)}{x-s}\bigg\}\,.
\end{equation}

At low energies, $M_0(s)$ and $M_2(s)$ can be matched to the chiral representation.
Namely in the limit of switching off the $\pi\pi$ FSI , i.e.\ $\Omega_l^0(s)\equiv 1$,
the polynomials $P_l^{n}(s)$ can be identified exactly with the expressions given in
Eq.~\eqref{eq.AmplitudeChiral}.
Therefore, the amplitudes including the FSI take the form
{\allowdisplaybreaks
\begin{align}
M_0(s)&=-\frac{2}{F_\pi^2}
\bigg[ c_1 \left(s-2m_\pi^2 \right)
+\frac{c_2}{2} \bigg(s+\vec{q}^2\Big(1  -\frac{\sigma_\pi^2}{3} \Big)\bigg)\bigg]\Omega_0^0(s)\,, \\
M_2(s)&=\frac{2}{3 F_\pi^2}c_2
\vec{q}^2\sigma_\pi^2 \Omega_2^0(s)
 \,. \label{eq.M02}
\end{align}
}%

Now we discuss the $\pi\pi$ phase shifts used in the calculation
of the Omn\`es functions. For the $S$-wave, we use the phase of the nonstrange pion
scalar form factor as determined in Ref.~\cite{Hoferichter:2012wf}, which yields a good description below the onset of the $K\bar K$ threshold. For the $D$-wave,
we employ the parametrization for $\delta_2^0$ given by the
Madrid--Krak\'ow collaboration~\cite{Pelaez}. Both phases are guided
smoothly to $\pi$ for $s\to\infty$.

It is then straightforward to calculate the $\pi\pi$ invariant mass spectrum and
helicity angular distribution for $\psi^\prime \to J/\psi
\pi^+\pi^-$ using
\begin{equation}
\frac{\diff\Gamma}{\diff \sqrt{s} \,\diff\cos\theta} =
\frac{\sqrt{s}\,\sigma_\pi |\vec{q}|}{128\pi^3 m_{\psi^\prime}^2}
\left|M_0(s)+M_2(s)
P_2(\cos\theta)\right|^2\,,\label{eq.pipimassdistribution}
\end{equation}
where the Legendre polynomial $P_2(\cos\theta)=(3\cos^2\theta-1)/2.$

\section{Phenomenological discussion} \label{pheno}

\begin{figure}
 \begin{center}
  \includegraphics[width=1.0\textwidth]{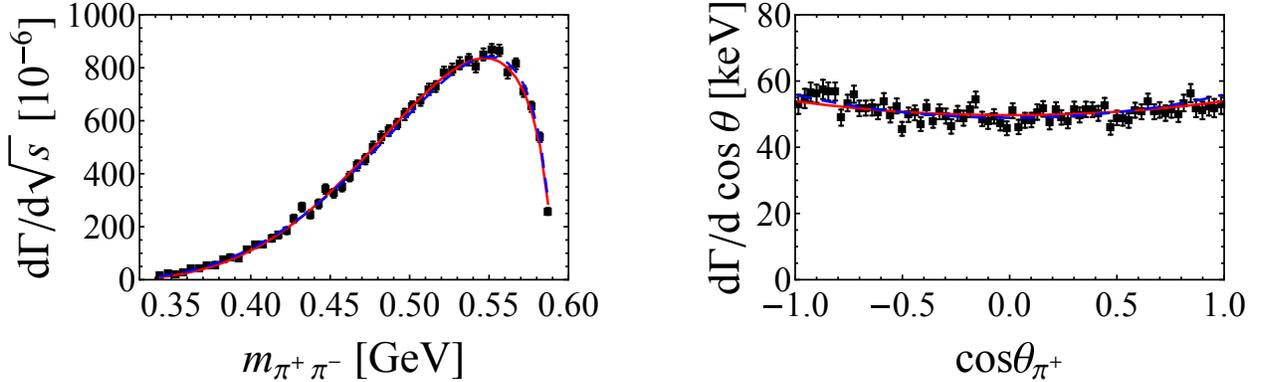}
 \end{center}
 \caption{Simultaneous fit to the $\pi\pi$ invariant mass distributions and the
 helicity angle distributions in $\psi^\prime \to J/\psi\pi^+\pi^-$. The
 red solid and blue dashed curves represent the theoretical fit results in the with $\pi\pi$ FSI and without
 $\pi\pi$ FSI cases, respectively. The data are taken from~\cite{Ablikim:2006bz}.
 }
 \label{fig:fitresults}
\end{figure}

The unknown parameters are the low-energy constants $c_1$ and $c_2$ in the
chiral Lagrangian~\eqref{LagrangianPsiPrimePsipipi}, which can be expressed
in terms of the chromopolarizability $\alpha_{\psi^\prime \psi}$ and the parameter $\kappa$ as in Eq.~\eqref{eq.Matching}.
In order to determine $\alpha_{\psi^\prime \psi}$ and $\kappa$, we fit the theoretical results to the experimental
$\pi^+\pi^-$ invariant  mass spectra and the helicity angular distribution from the BES $\psi^\prime \to J/\psi\pi^+\pi^-$
decay data~\cite{Ablikim:2006bz}, and the corresponding decay width $\Gamma(\psi^\prime \to J/\psi\pi^+\pi^-)$~\cite{Tanabashi:2018oca}. The fit results are plotted in Fig.~\ref{fig:fitresults}, where the red solid and
blue dashed curves represent the results with or without the $\pi\pi$ FSI, respectively. The fit parameters as well as the
$\chi^2/\text{d.o.f.}$ are shown in Table~\ref{tablepar}. One observes that the experimental data can be well described regardless
of whether the FSI is included. This is due to the simple shapes of the $\pi\pi$ invariant mass distribution and the helicity angular distribution in this process and does not mean the FSI is not important. Since the dipion mass invariant mass reaches about 600 MeV in such a decay, the $\pi\pi$ FSI is known to be strong in this energy range and needs to be considered. On the other hand, one can readily see from Eqs.~\eqref{eq.Matching} and~\eqref{eq.AmplitudeChiral}, while the chromopolarizability $\alpha_{\psi^\prime \psi}$ determines the overall decay rate, the parameter $\kappa$ characterizes the $D$-wave contribution, and we do not find significant correlation between $\alpha_{\psi^\prime \psi}$ and $\kappa$. 

\begin{table}
\caption{\label{tablepar} The parameter results from the fits of
the $\psi^\prime \to \psi \pi\pi $ processes with and without the $\pi\pi$ FSI.}
\renewcommand{\arraystretch}{1.2}
\begin{center}
\begin{tabular}{lcc}
\toprule
         & Without $\pi\pi$ FSI
         & With $\pi\pi$ FSI\\
\hline
$|\alpha_{\psi^\prime\psi}| $ (GeV$^{-3}$)   &   $ 2.37\pm 0.02$           & $1.44\pm 0.02 $  \\
$\kappa$   &   $ 0.135\pm 0.005$ & $0.139\pm 0.005 $   \\
\hline
 $\frac{\chi^2}{\rm d.o.f}$ &  $\frac{115.3}{120-2}=0.98$   &
$\frac{117.6}{120-2}=1.00$  \\
\botrule
\end{tabular}
\end{center}
\renewcommand{\arraystretch}{1.0}
\end{table}

We observe that the $\pi\pi$ FSI modifies the value of the chromopolarizability $\alpha_{\psi^\prime \psi}$ significantly, and resultant value with the FSI is almost 1/2 of that without the FSI. The obtained value
with the FSI, $|\alpha_{\psi^\prime \psi}|=(1.44\pm 0.02)$ GeV$^{-3}$, coincides with the suspicion $\alpha_{J/\psi}\geq |\alpha_{\psi^\prime \psi}|$~\cite{Sibirtsev:2005ex} with the value $\alpha_{J/\psi}=(1.6\pm 0.8)$ GeV$^{-3}$ from the calculation~\cite{Polyakov:2018aey} based on the recent lattice QCD data of $J/\psi$-nucleon potential~\cite{Sugiura:2017vks}.
It should be mentioned that the value of $\alpha_{\psi^\prime \psi}$ with the FSI obtained here is different from the one in Ref.~\cite{Guo:2006ya},
$|\alpha_{\psi^\prime \psi}|=(0.83\pm 0.01)$ GeV$^{-3}$,
and also our result without the FSI slightly differs from those in
Refs.~\cite{Voloshin:2004un,Guo:2006ya}. The reasons are that the chiral unitary approach instead of dispersion theory is used to account for the FSI in~\cite{Guo:2006ya}, and we use the updated experimental data~\cite{Ablikim:2006bz,Tanabashi:2018oca} and
a general theoretical amplitude rather than the one only containing the $S$-wave as employed in Refs~\cite{Voloshin:2004un,Guo:2006ya}.

For the parameter $\kappa$, as shown in Table~\ref{tablepar} its value is affected little by the $\pi\pi$ FSI.
One notes that a detailed study of the $\psi^\prime \to J/\psi\pi^+\pi^-$ process using the Novikov-Shifman model~\cite{Novikov:1980fa} have been performed by BES~\cite{Bai:1999mj}, and based on the joint $m_{\pi^+\pi^-}$-$\cos\theta_{\pi^+}$ distribution this parameter was determined as $\kappa=0.183\pm0.002\pm0.003$. We have tried fitting the same old BES data~\cite{Bai:1999mj}, and our $\kappa$ changes slightly and is still much smaller than the BES one. In the Novikov-Shifman model, the $\psi^\prime \to J/\psi\pi^+\pi^-$ amplitude reads~\cite{Novikov:1980fa}
\begin{align}\label{eq.AmplitudeShifman}
M \propto
\bigg\{s-\kappa\left(m_{\psi^\prime}-m_\psi\right)^2\left(1+\frac{2m_\pi^2}{s}\right)+\frac{3}{2}\kappa\bigg[(m_{\psi^\prime}-m_\psi)^2-s\bigg]
\sigma_\pi^2\left(\cos^2\theta-\frac{1}{3}\right)
\bigg\}\,.
\end{align}
If we make the same approximation, namely neglect the $O(m_\pi^2)$ terms except the $m_\pi^2/s$ ones, as in Ref.~\cite{Novikov:1980fa} and
set $(m_{\psi^\prime}+m_\psi)^2-s \approx (m_{\psi^\prime}+m_\psi)^2$ in the expression of 3-momentum $\vec{q}$, our amplitude without the $\pi\pi$ FSI agrees with Eq.~\eqref{eq.AmplitudeShifman}. While numerically we find that some neglected $O(m_\pi^2)$ terms
are at the same order as the $\kappa\left(m_{\psi^\prime}-m_\psi\right)^2$ term in Eq.~\eqref{eq.AmplitudeShifman}, and this may account for the difference of $\kappa$ between ours and that in Ref.~\cite{Bai:1999mj}. On the other hand, we have checked that the contribution of the $D$-wave, which is characterized by the parameter $\kappa$, to the total rate is less than two percent, and the same observation has been made in~\cite{Bai:1999mj}.

\section{Conclusions}
\label{conclu}

We have used dispersion theory to study the $\pi\pi$ FSI in the decay
$\psi^\prime \to J/\psi \pi^+\pi^-$. Through fitting the data of the $\pi\pi$ mass spectra and
the angular $\cos\theta$ distributions, the values of the chromopolarizability $\alpha_{\psi^\prime \psi}$
and the parameter $\kappa$ are determined. It is found that the effect of the $\pi\pi$ FSI is quite sizeable in the
chromopolarizability $\alpha_{\psi^\prime \psi}$, and the one with FSI is almost $1/2$ of that without
FSI. While the parameter $\kappa$, which accounts for the $D$-wave contribution, is affected little by the $\pi\pi$ FSI.
The results obtained in this work would be valuable to understand the chromopolarizability of charmonia, and will have applications for the
studies of the nucleon-charmonia interaction.

\section*{Acknowledgments}

We acknowledge Feng-Kun Guo for the proposal for this work and for the useful comments on
the manuscript. This research is supported in part by the Fundamental
Research Funds for the Central Universities under Grant
No.~06500077.

\end{document}